\documentclass[aps,twocolumn,showpacs, prl]{revtex4}
\usepackage{amsmath,amsfonts,amssymb,graphics,graphicx,epsfig,color,times,bbm}
\usepackage{amsbsy,latexsym}
\usepackage{amsfonts}
\usepackage{amssymb}
\usepackage[mathscr]{eucal}
\newcommand{\dg}{^\dagger}

\newcommand{\be}{\begin{equation}}
\newcommand{\ee}{\end{equation}}

\newcommand{\ex}[1]{\mathrm{e}^{#1}}
\newcommand{\id}{\openone}

\newcommand{\tr}{{\rm tr}}

\newcommand{\ket}[1]{\left|{#1}\right\rangle}
\newcommand{\bra}[1]{\left\langle{#1}\right|}
\newcommand{\braket}[2]{\langle{#1}|{#2}\rangle}
\newcommand{\ketbrad}[1]{\left|{#1}\rangle\!\langle{#1}\right|}

\newcommand{\ea}{\emph{et al.}}

\begin{document}

\title{Phase-Covariant Quantum Benchmarks}

\author{J.~Calsamiglia}
\affiliation{Grup de F\'{\i}sica Te\`{o}rica, Universitat
Aut\`{o}noma de Barcelona, 08193 Bellaterra (Barcelona), Spain}
\author{M. Aspachs}
\affiliation{Grup de F\'{\i}sica Te\`{o}rica, Universitat
Aut\`{o}noma de Barcelona, 08193 Bellaterra (Barcelona), Spain}
\author{R.~Mu\~{n}oz-Tapia}
\affiliation{Grup de F\'{\i}sica Te\`{o}rica, Universitat
Aut\`{o}noma de Barcelona, 08193 Bellaterra (Barcelona), Spain}
\author{E. Bagan}
\affiliation{Grup de F\'{\i}sica Te\`{o}rica, Universitat
Aut\`{o}noma de Barcelona, 08193 Bellaterra (Barcelona), Spain}


\begin{abstract}
We give a quantum benchmark for teleportation and quantum storage experiments suited for pure and mixed test states. The benchmark is based on the average fidelity over a family of phase-covariant states 
and certifies that an experiment can not be emulated by a classical setup, i.e., by a measure-and-prepare scheme. We give an analytical solution for qubits, which shows important differences with standard state estimation approach, and compute the value of the benchmark for coherent and squeezed states, both pure and mixed.  

\end{abstract}
\pacs{03.67.Hk, 03.65.Ta}

\maketitle
\emph{Introduction:}\label{intro}
A central question in quantum information theory is whether a particular
quantum protocol can be realized with the same
efficiency by classical means; ultimately, quantum information stands on the advantage of quantum
over classical systems in performing certain tasks. This question is relevant in experimental
implementations of such protocols, as they become imperfect in real, unavoidably noisy, experiments.
It is then essential to assess whether the same experimental result
could have been obtained by using only classical (less costly) resources. 
So, for example,
in a particular teleportation experiment ---which involves
generating entanglement, performing complicated Bell measurements, fighting
decoherence, etc.--- one may ask whether  the
same goal, i.e., mapping the state of a system onto a second system at
a different space-time location, could have been achieved by measuring the quantum state of the first system, transmitting the
collected information, and preparing
the state of the second accordingly.  In this letter, we propose and calculate
quantum benchmarks that certify that a certain implementation of teleportation cannot be realized classically, or  more precisely, by a measure-and-prepare strategy.
Of course, these benchmarks also apply to any protocol that can be understood as a realization of the identity channel, e.g., quantum storage.
Our benchmarks are based
on phase-covariant families of states, and thus they are computationally manageable
and experimentally feasible (shifting phases is straightforward in, e.g., optical experiments). Last but not least, our benchmarks  apply when test states are mixed, which is the standard situation in experiments.
Previously proposed quantum  benchmarks~\cite{hammerer_quantum_2005,serafini_teleportation_2007,adesso_quantum_2008} were either restricted to pure states or were not strict bounds, since they were based on sub-optimal classical strategies (see footenote 15 in \cite{adesso_quantum_2008}).
We give rigorous quantum benchmarks for both pure and mixed Gaussian states that can be immediately  applied to current experiments on continuos-variable (CV) systems, such as optical fields and  atomic ensembles \cite{furusawa_unconditional_1998,julsgaard_experimental_2004}.

\emph{Background and methods:}
The literature on quantum benchmarks originated in the context of  CV  teleportation experiments, the first of which was performed in~1998 by A.~Furusawa 
\emph{et al.}~\cite{furusawa_unconditional_1998}. In this experiment, optical coherent states were teleported using squeezed-state entanglement.  
As benchmark, they used the average fidelity,~${\cal F}$, that can be attained without entanglement (with the EPR beams replaced by the vacuum). 
Braunstein \emph{et al.}~\cite{braunstein_criteria_2000} proposed a more rigorous benchmark for CV teleportation. They considered the fidelity 
between an input state $\ket{\psi_{\rm in}}$ and the corresponding state, $\rho_{\rm out}$, outputted from a measure-and-prepare channel,  i.e., $\rho_{\rm out}=\sum_{\chi}p(\chi|\psi_{\rm in})\rho_{\chi}$, 
where $\rho_{\chi}$ is the reconstructed or guessed state based on outcome~$\chi$ of the measurement and $p(\chi|\psi_{\rm in})$ is the conditional probability of obtaining~$\chi$ given that the signal state is~$|\psi_{\rm in}\rangle$. Their benchmark is given by the maximum fidelity averaged over a conveniently chosen set~$\Omega$ of input states.
It should be stressed  that the choice of~$\Omega$ plays a significant role. It should contain necessarily more than one state (otherwise the test is passed by a trivial classical strategy). The maximum fidelity tends to decrease with the size of the test family. When the input states are drawn from a whole $d$-dimensional Hilbert space, the optimal fidelity is known to be
$
\mathcal{F}=2/(d+1)
$ \cite{bruss_optimal_1999}.
In the
case of CV systems, $d\rightarrow\!\infty$,  this would mean that any non-zero fidelity would certify that quantum resources are being used. However, it is utterly unrealistic to assume that one can test the channel with such a large family of input states. 
In order to have realistic and practical thresholds, Braunstein \emph{et al.} \cite{braunstein_criteria_2000}  chose~$\Omega$ to be the set of coherent states with normally distributed amplitude around~$|\alpha|=0$ and a fixed given variance,
and gave a classical strategy that recently has been proved to be optimal  by Hammerer \ea~\cite{hammerer_quantum_2005}.

Adesso and Chiribella~\cite{adesso_quantum_2008} have very recently
proposed a quantum benchmark  suited pure single-mode squeezed states and derived upper and lower
bounds for~squeezed thermal states. Their quantum benchmark is taken to be the maximum  averaged fidelity over all the outcomes~$\chi$  of an optimal measurement and over an ensemble~$\Omega$ of input states $\rho_{\rm in}(r)$ whose squeezing parameter~$r$ is distributed with a given prior probability (in \cite{serafini_teleportation_2007} a benchmark for pure squeezed states drawn from a microcanonical ensemble was considered). In this approach, (i) the output state is assumed to belong to  the input family, i.e., is of the form~$\rho(r_\chi)$, and (ii) the ``verifier''  has to know what outcome $\chi$ has been obtained at every single use of the hypothetical measure-and-prepare channel. This makes the so-defined fidelity difficult to compare with the fidelity of the real channel, which does not even have to involve measurements.  As to (i), 
we will see that this choice of output state is typically sub-optimal and, hence, the corresponding fidelity might not be a strict quantum benchmark, as the authors noticed.

Here we propose a benchmark based on a general phase-covariant family (pure or mixed):
\begin{equation}
 \mathcal{F}=\int \frac{d\phi}{2\pi} F(\rho_{\phi},\rho_\mathrm{av}^{\phi}),
 \label{eq:defbench}
\end{equation}
where $F(\rho_{1},\rho_{2})=(\tr\sqrt{\sqrt{\rho_{1}}\rho_{2}\sqrt{\rho_{1}}}\,)^{2}=
(\tr |\sqrt{\rho_{1}}\sqrt{\rho_{2}} | )^2$ is the fidelity, $\rho_\phi$ is the input test state  defined as  $\rho_{\phi}=U(\phi)\rho_{0}U(\phi)\dg$ with $U(\phi)=\sum_{n}\ex{i\phi n}\ketbrad{n}$ ($|n\rangle$ are Fock states), and 
$\rho_{\mathrm{av}}(\phi)=\sum_{\chi}p(\chi|\rho_{\phi})\rho_{\chi}$ 
is the state outputted by the measure-and-prepare channel. In contrast to (i)~above~\cite{adesso_quantum_2008,serafini_teleportation_2007}, guessed states $\rho_{\chi}$ are not constrained to belong to the input-state family. 
In contrast to~(ii),  $\rho_{\rm av}$ is the output of the channel, with no reference to the way it has been implemented or to any further information. Since the fidelity for mixed states is non-linear Adesso and Chiribella's benchmark and ours will typically differ. From the convexity of the fidelity it follows that 
theirs is a lowerbound to~$ \mathcal{F}$ and hence it 
 is arguably not a proper quantum benchmark.
 A benchmark based on phase-covariant states is appealing from an experimental point of view, since phases are easy to vary without affecting other relevant parameters (e.g., in the presence of losses,  varying the degree  of squeezing leads to a change in purity of the test states, which is very hard to compensate).

\medskip

Given a strategy characterized by POVM elements $\{O_{\chi}=\ketbrad{\xi_\chi}\}$ (they can be taken to be of rank one without loss of generality and $|\xi_\chi\rangle$ are not normalized), and
corresponding guesses $\{\rho_\chi\}$, one  defines a covariant strategy by 
$\{O_{\chi,\theta}=1/(2\pi)\, U_\theta O_{\chi} U_\theta^\dagger\}$ and
$\{\rho_{\chi,\theta}=U_\theta\rho_\chi U_\theta^\dagger\}$.
One can show that the optimal strategy can always be chosen to be covariant
\footnote{Starting from the definition \eqref{eq:defbench} and using the the properties:  (i) for any two unitaries $U$ and $V$ the trace norm satisfies $\tr|U B V|=\tr |B|$; (ii) the fidelity is convex, and (iii) $U(\phi)U(\theta)\dg=U(\phi-\theta)$.}  and 
\begin{equation}
 \mathcal{F}\!=\! \left( \tr\! \left| \sqrt{\rho_0} \sqrt{\rho_{\mathrm{av}}}\right|\right)^2
 \!;\
 \rho_\mathrm{av}\!=\!\!\int\!d\theta \sum_\chi p(\chi,\theta |\rho_{0}) \rho_{\chi,\theta}.
\label{eq:sym}
\end{equation}
Note that aside from the group parameter~$\theta$, one needs to specify also the ``seed'' for both POVM ($O_\chi$) and guess states~($\rho_\chi$). For single-seed strategies, the completeness relation fixes the POVM, which turns out to correspond to
the \emph{phase-measurement}~\cite{holevo_probabilistic_1982}:~$\ket{\xi}= \sum_n \ket{n}$ (up to some arbitrary phases).

The classical fidelity \eqref{eq:sym} can be conveniently written as
\begin{equation}
 \mathcal{F}=
\max_{K}
\left(\tr_B\sqrt{\tr_A \sqrt{\rho_0}\otimes\sqrt{\rho_0} K
\sqrt{\rho_0}\otimes\sqrt{\rho_0}}\right)^2,
 \label{eq:opt}
\end{equation}
where $\tr_A$ (similarly $\tr_B$) stands for the obvious partial trace and
$
 K=\int\!d\theta\sum_\chi O_{\chi,\theta}\otimes
\rho_{\chi,\theta}.
$
Optimizing the classical strategy amounts to maximizing the trace squared in~(\ref{eq:opt}) 
over the set of positive operators acting on~$\mathcal{H}\otimes\mathcal{H}$ that are separable, invariant under
bilateral transformations~$U_\theta\otimes U_\theta$, and that
fulfill  $\tr_B K
= \openone_A$.
For pure states, Eq.~(\ref{eq:opt}) can be simplified to give $
 \mathcal{F}=\bra{\psi_{0}}\!\bra{\psi_{0}}K\ket{\psi_{0}}\!\ket{\psi_{0}}
$, with $\rho_{0} =\ketbrad{\psi_0}$.
This leads to the pure-state estimation approach  introduced in~\cite{navascues_pure_2008}.
For a given POVM with seeds $\{\ket{\xi_\chi}\}$  the optimal fidelity can also be written as,
\be
\mathcal{F}=\sum_{\chi}\sup_{\psi_\chi}\bra{\psi_{\chi}}A_{\chi}\ket{\psi_{\chi}}
=\sum_{\chi}\left\| {A_{\chi}}\right\|_{\infty} ,
\label{eq:clfid}
\ee
where 
$A_{\chi}=\int d\phi/(2\pi) |\langle\xi_\chi|\psi_\phi\rangle|^2 |\psi_\phi\rangle\langle\psi_\phi|$.
Thus, if the POVM is fixed, the maximum fidelity is given by  the largest eigenvalue of the operator $A_{\chi}$ and the optimal guess seed,~$|\psi_\chi\rangle$, by the corresponding normalized eigenvector~\cite{hammerer_quantum_2005}.

If one restricts the guess-states to be in~$\Omega$ (as done in ~\cite{adesso_quantum_2008, serafini_teleportation_2007}), things simplify considerably, specially for pure states where no optimization is required since the optimal POVM is known to be the phase-measurement \cite{holevo_probabilistic_1982}.
In our case,  no assumption about the optimal POVM nor about  the guess is made and we have to resort to more powerful techniques.

Semi-definite programming (SDP) \cite{vandenberghe_semidefinite_1996} is an area of convex optimization that was developed in the last decade and that has recently found several applications in the field of Quantum Information~ \cite{doherty_complete_2004, SDPQI,navascues_pure_2008}. Its aim  is to minimize a linear objective function subject to semidefiniteness constraints involving symmetric matrices that are affine in the variables:%
\be
\min_{\mbox{\small\boldmath$x$}} \mbox{\boldmath$c$}^T \mbox{\boldmath$x$}  \; \mbox{ subject to } \;
F(\mbox{\boldmath$x$})=F_{0}+\sum_{i}x_{i}F_{i}\geq 0 
\label{eq:SDPprim},
\ee
where $F_{i}$ are hermitian matrices of arbitrary dimension, and the inequality means that $F(\mbox{\boldmath$x$})$ is positive semidefinite. 
There are a number of freely available software packages to solve SDP problems. In this work we have used the {YALMIP} matlab toolbox~\cite{lfberg__2004} together with the {SDPT3} solver~\cite{toh_sdpt3_1999}.

For pure input states, the maximization in~(\ref{eq:opt}) can be immediately cast into a SDP problem, by writing the matrices $K$ and $\rho_{0}$ in a basis of Hermitian matrices. Whereas the positivity and bilateral invariance constrains are already in the desired form, the  separability condition need be imposed through a hierarchy of constrains based on PPT symmetric extensions~\cite{doherty_complete_2004}. In this work we will stick to the lowest level of this hierarchy, imposing only positivity under partial transposition (PPT), i.e.,  $K^{\Gamma}\geq 0$. Since  positive partial transposition provides a necessary, but in general not sufficient, condition for separability, the resulting optimal value, $\mathcal{F}^\Gamma$, gives an upper bound to $\mathcal{F}$, 
and hence still provides a valid quantum benchmark.

When the input test states are mixed, the objective function in~(\ref{eq:opt}) becomes truly non-linear and  the optimization problem does not immediately fall into the SDP category. However, one can make use of Uhlmann's theorem~\cite{uhlmann_transition_1985} and recast~${\cal F}$ in~(\ref{eq:sym}) as
$
\mathcal{F}=\max_{\Psi_{\mathrm av}} |\braket{\Psi_{0}}{\Psi_{\mathrm av}}|^2=
\min_{\sigma_{\mathrm{av}}} (-\bra{\Psi_{0}}\sigma_{\mathrm{av}}\ket{\Psi_{0}})
$,
where $\ket{\Psi_{0}}$ and $\sigma_{\rm av}=\ket{\Psi_\mathrm{av}}\bra{\Psi_\mathrm{av}}$ are purifications of $\rho_0$ and $\rho_{\rm av}$ respectively.  Without loss of generality, the purity condition on $\sigma_{\rm av}$ can be lifted. 
With this, 
the objective function becomes a linear function of the optimization variables. 
Constraints are also of the SDP form in (\ref{eq:SDPprim}): 
(i)~$\tr_{B} \sigma_{\mathrm{av}}=\rho_{\rm av}=\tr_{A}(\rho_{0}\otimes \id\, K)$; 
(ii) $ \sigma_{\mathrm{av}}\geq 0$ and~$\tr\,\sigma_{\mathrm{av}}=1$; and
(iii) the same conditions on~$K$ as above.

\emph{Qubit states:}
It will become apparent as we proceed, that analytical solutions to the benchmark problem for general mixed states are exceedingly hard to obtain.
A remarkable exception are qubit states, which we
discuss next. 

The input-state
family is defined by $\rho_\phi=U_{z,\phi}\rho_0U_{z,\phi}^\dagger$, where $U_{z,\phi}$ is a rotation of an angle $\phi$ around the $z$-axis (similarly for $U_{y,\theta}$, etc.), and the seed input state is
\mbox{$\rho_0=U_{y,\theta}\{p |\kern-.35em\uparrow\rangle\langle\uparrow\kern-.35em|+(1-p) |\kern-.35em\downarrow\rangle\langle\downarrow\kern-.35em| \}U_{y,\theta}^\dagger$}, for fixed azimuthal angle~$\theta$ and probability~$p\equiv (1+\eta)/2$ (so that $\eta$ is the modulus of the Bloch vector).
To calculate the quantum
benchmark for this family of qubit states, we use~\eqref{eq:opt}.  In the up/down basis
the most general~$K$ can be written as~$K={\rm blockdiag}(a,B,c)$,
where $B$ is a $2\times2$ positive semi-definite matrix
and $a$, $c$ are non-negative numbers.
For ${\mathbb C}^2\otimes{\mathbb C}^2$ the partial transposition criteria provides a
necessary and sufficient condition for separability, i.e., $K$ is separable
$\Leftrightarrow K^{\Gamma}\geq 0 \Leftrightarrow a c- |B_{12}|^2\geq0$. Finally, the
condition $\tr_B K =\openone_{A}$ implies \mbox{$a+B_{11}=c+B_{22}=1$.} A~tedious but straightforward calculation leads to the optimal $K$ (that maximizes ${\cal F}$):
$a=\cos^2(\zeta/2)$, $c=\sin^2(\zeta/2)$, $B_{12}=B_{21}=\sqrt{a c}$,
%
%
where the (azimuthal) angle~$\zeta$ is
\begin{equation}
\zeta=\arctan{\eta^2\sin^2\theta+\sqrt{(1-\eta^2)(4-\eta^2\sin^2\theta)}\over2\eta\cos\theta} .
\end{equation}
It is
a simple exercise to check that these values of~$a$, $B$ and~$c$ correspond to the phase-covariant POVM $\{1/\pi
U_{z,\phi}U_{y,{\pi\over2}}\ketbrad{\uparrow}U_{y,{\pi\over2}}^\dagger U_{z,\phi}^\dagger\}$, with the associated guess
$|\phi\rangle=U_{z,\phi}U_{y,\zeta}|\kern-.35em\uparrow\rangle$. The
resulting  benchmark for qubit states can be cast as
\begin{equation}
\mathcal{F}=\frac{1}{2} \left(
1+\eta{\cos\theta\over\cos\zeta}\right) \label{eq:fclqubits}.
\end{equation}
Some remarks are in order: (i) for any outcome $\phi$ the corresponding guess is a pure state and, hence,  (ii) it  does not belong to the original family; moreover, (iii) its Bloch vector is not proportional to that of the signal state ($\zeta\not=\theta$, unless $\theta=\pi/2$). In quantum state estimation it is usually assumed 
that the guess after a measurement is one of the possible input states. Our result show that this assumption is not always well-founded.
(iv) Recalling the definition of $K$ right after Eq.~(\ref{eq:opt}), we see that only one value of $\chi$ (a single  ``seed") is required for qubits. Property (iv) is specific of ${\mathbb C}^2$. For ${\mathbb C}^4$ one can already find examples where the phase-measurement is not optimal.

We now move to the benchmarks for CV gaussian states. We consider displaced and squeezed thermal states that are obtained by the action of the displacement, $D(\alpha)=\exp[\alpha (a\dg-a)]$, and the squeezing, $S(r)=\exp[\frac{r}{2}(a^2-a^{\dagger 2})]$, operators over a thermal state $\rho_{\beta}\!=\!(1\!-\!\ex{-\beta})\ex{-\beta\,a\dg\! a}$ of purity $\mu=\tanh\beta/2$ (with $\alpha>0$ and $r>0$).

\begin{figure}[htp]
\begin{picture}(230,80)
\put(0,77){{\scriptsize $\mathcal{F}$}}
\put(125,77){{\scriptsize $\mathcal{F}$}}
\put(50,-6){{\scriptsize $ |\alpha|^2$}}
\put(180,-6){{\scriptsize $\lambda$}}
\put(-10,0){\includegraphics[width=1.65in]{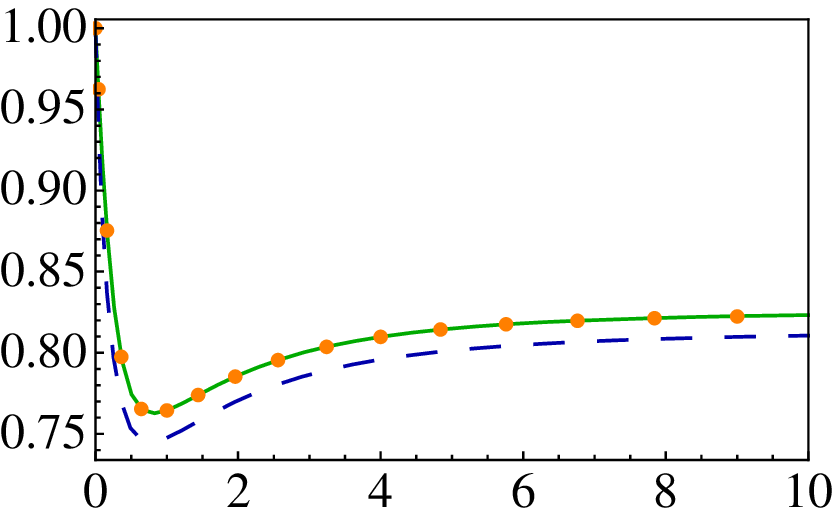}}
\put(120,0){\includegraphics[width=1.6in]{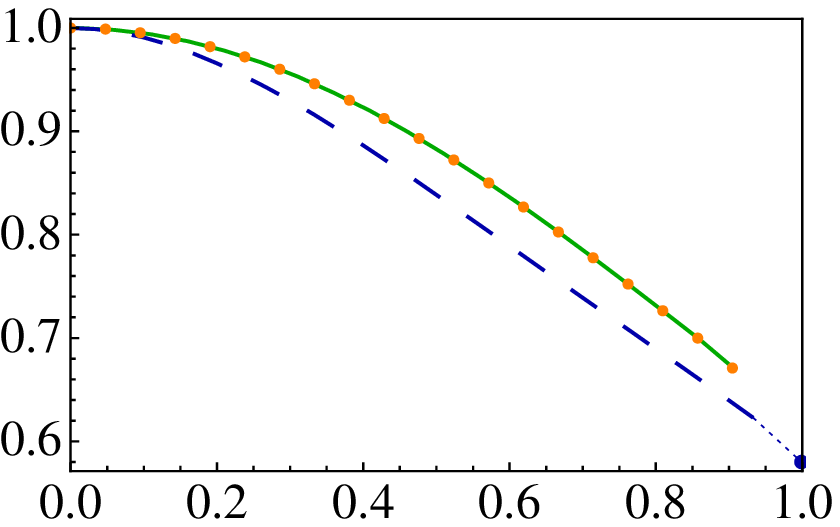}}
\end{picture}
\caption{Plot of the average fidelity for pure coherent states (left) and pure squeezed states (right). Dots: bound obtained imposing $K^\Gamma\geq0$ (see main text).
Solid: phase-measurement and optimal guess.
Dashed: phase-measurement and guess from input family. Dotted:  extrapolation of dashed line to infinite squeezing,~$\lambda=1$.}\label{fig:bestbenchcoh}
\end{figure}

\emph{CV pure states:}
We start by computing the bound  $\mathcal{F}^{\Gamma}$ using the SDP approach for coherent states.
For this purpose we use a truncated Fock basis
and approximate low amplitude coherent states by $\ket{\alpha}\approx \ex{{-\alpha^2/2}}\sum_{n=0}^{N}\alpha^n/\sqrt{n!}\ket{n}$.  Figure  \ref{fig:bestbenchcoh} (dots) shows the results for coherent states with mean photon number between zero and $\alpha^2=10$ (the truncation error within this range of $\alpha$  and $N=23$ is always lower than $10^{-4}$). 
In addition we can calculate the optimal fidelity when restricting to the single-seed covariant POVM. With this choice of POVM the problem reduces to calculating the maximum eigenvalue of the matrix in  \eqref{eq:clfid},
$
A=
\ex{-2\left|\alpha\right|^{2}}\sum_{n}{\alpha^{2n}}/{n!}
\ket{\phi_n}\!\bra{\phi_n}
$, 
where  $\ket{\phi_n}=\sum_{l=0}^{n}\sqrt{\binom{n}{l}}\ket{l}$.
The values for different input intensities are shown in Fig. \ref{fig:bestbenchcoh} (solid line) and agree with those obtained from the SDP optimization. This indicates that the benchmark given by $\mathcal{F}^{\Gamma}$ is attainable with the phase-measurement (at least within the precision of our numerical analysis).
We note that the eigenvector of $A$ with largest eigenvalue, which is the optimal guess,
resembles a coherent state but  is strictly different. The dashed line in Fig. \ref{fig:bestbenchcoh} is the fidelity obtained when the guess is forced to be a coherent state and, although it has a similar behavior, it shows a clear gap  with the optimal bound. In this case no optimization is required and $\mathcal{F}$ can be obtained by numerical integration of
\be\label{eq:f-coh}
\mathcal{F}=\int \frac{d\phi}{2\pi}|\braket{\xi}{\alpha \ex{i \phi}}|^2|\braket{\alpha}{\alpha \ex{i \phi}}|^2.
\ee
From the above equation it is possible to find the analytical value of $\mathcal{F}$ in the limit $\alpha\to \infty$. In this regime, the outcome
probabilities can be approximated by $|\braket{\xi}{\alpha \ex{i
\phi}}|^2\simeq \sqrt{2\alpha^2/\pi}\exp[-2\alpha^2\phi^2]$, where
we have used the Gaussian limiting expression of a Poisson
distribution, $|\braket{\alpha}{\alpha \ex{i
\phi}}|^2\approx\exp[-\alpha^2\phi^2]$, and  extend  the range of integration to $(-\infty,\infty)$ to find
\mbox{$\mathcal{F}_{\alpha\to\infty}=\sqrt{2/3}$}.

The analytical value of $\mathcal{F}$ in the large $\alpha$ limit can also be obtained in the general case of unrestricted guess states.
In this case one has to calculate  the maximum eigenvalue of $A$ defined in
Eq.~\eqref{eq:clfid}, which can be done by calculating the limit
$p\to\infty$ of its p-norm $||A||_p=(\tr A^p)^{1/p}$. We have
\begin{equation}\label{eq:p-norm}
(||A||_p)^p=\tr A^p=\int\prod_{j=1}^{p} d\phi_j \,p(\chi | \phi_j)
\braket{\alpha_j}{\alpha_{j+1}},
\end{equation}
where  $\alpha_{p+1}\equiv \alpha_1$. Using the above
approximation on the outcome probabilities and
$\braket{\alpha_i}{\alpha_{j}} \approx \exp\{-\alpha^2 [ i (\phi_i-\phi_j)
+1/2(\phi_i-\phi_j)^2 ]\}$,  we obtain
\begin{equation}\label{eq:p-norm-coh-2}
\tr A^p \simeq \left(\frac{2 \alpha^2}{\pi}\right) ^{p/2}\!\!\int
d^p{\phi}\, \ex{-\frac{\alpha^2}{2} \phi^t \cdot C_p\cdot \phi} =\!
\frac{2^{p}}{\sqrt{\det C_{p}}},
\end{equation}
where  $C_p$ is a symmetric matrix with elements $[C_p]_{ij}=6\delta_{ij}-\delta_{i+1,j}-\delta_{i,1}\delta_{j,p}$ for $i\leq j$.
It is convenient to write  $ C_p=M_p+2a\openone_p$,
with $a=3$. Then, 
$\det (M_p+2a\openone_p)=Q_p(a)$ is a (characteristic)
polynomial in $a$. It easy to check
that $Q_p(a)=2[T_p(a)-1]$, where
$T_p(a)=[(a+\sqrt{a^2-1})^p+(a-\sqrt{a^2-1})^p]/2$ are the Chebyshev
polynomials of the first kind. In the limit $p\!\to\!\infty$ the second term in $T_p(a)$ becomes negligible, hence
$Q_{p}(3)\approx (3+\sqrt{8})^p=(\sqrt{2}-1)^{2p}$, 
and
\begin{equation}
\mathcal{F}=\lim_{p \rightarrow \infty} 2 (\det
C_p)^{1/2p} =
2(\sqrt{2}-1)\approx 0.8284,
\end{equation}
which indicates that the difference between the fidelities corresponding to restricted (guess in $\Omega$) and unrestricted (general guess) strategies  persists also in the asymptotic regime.

For pure squeezed states we proceed as before by working in a truncated basis:
$
\ket{\lambda}=(1\!\!-\lambda^{2})^{1/4}\sum_{n=0}^{N}\left(-{\lambda}/{2}\right)^{n}
{\sqrt{(2n)!}}/{n!}\ket{2n}
$.
Figure~\ref{fig:bestbenchcoh} shows the SDP results, together with the bound obtained from the phase-measurement with optimal guess (the maximum eigenvalue of $A$), and  from the phase-measurement with (restricted) squeezed guess. The latter is obtained by numerically integrating
\begin{equation}\label{eq:fidsqsubopt}
\mathcal{F}\!=\!\int \kern-0.5em \frac{d\phi\,(1-\lambda^{2})^{2}}{\sqrt{1+\frac{4\lambda^{2}}
{(1\!-\!\lambda^{2})^{2}}\!\sin^{2}\!{\phi}}}\left|\sum_{n}\frac{\sqrt{\!(2n)!}}{n!}\left(\frac{\lambda \ex{2 i\phi}}{2}\right)^{n}\right|^{2}.
\end{equation}

The SDP bound and the phase-measurement fidelity agree within numerical precision. As for coherent states,  restricting the guess states be in $\Omega$ lowers the bound substantially.  The latter bound can again be computed in the limit $\lambda\to 1$ by noticing that the
dominant behavior of the sum in Eq.~\eqref{eq:fidsqsubopt} is dictated by the
large $n$ terms~\cite{bagan_phase_2008}. Using the
Stirling's approximation,
the sum in~\eqref{eq:fidsqsubopt} can be written as a polylogarithm function $\mathrm{Li}_{1/4}(\lambda \ex{2i\phi})/\pi^{1/4}$. Taking into account that the dominant contribution to the integral comes from the region $\phi\sim 0$, we get
$\mathcal{F}\simeq 0.58$. 

\begin{figure}[!htp]
\begin{picture}(230,80)
\put(0,77){{\scriptsize $\mathcal{F}$}}
\put(125,77){{\scriptsize $\mathcal{F}$}}
\put(50,-6){{\scriptsize $ \alpha^2$}}
\put(180,-6){{\scriptsize $r$}}
\put(-10,0){\includegraphics[width=1.62in]{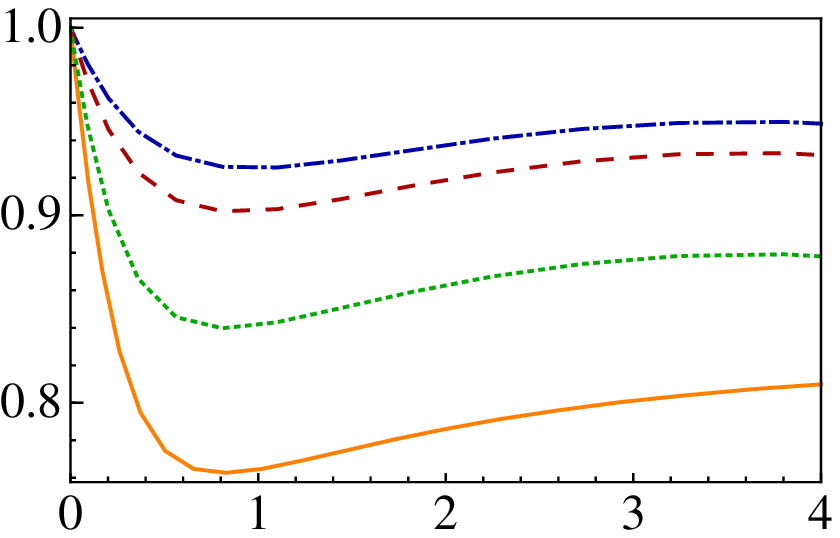}}
\put(120,0){\includegraphics[width=1.65in]{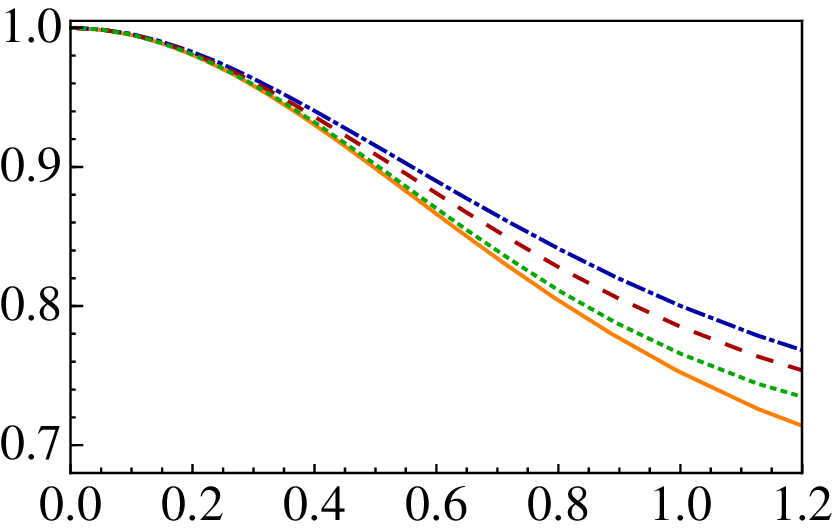}}
\end{picture}
\caption{Quantum benchmarks for displaced thermal states (left) and squeezed thermal states (right) for different purities: $\mu=1$ (solid line),
$\mu=0.95$ (dots), $\mu=0.8$ (dash) and $\mu=0.7$ (dash-dot).}
\label{fig:bestbenchmix}
\end{figure}

\emph{CV mixed states:} The case of mixed states is remarkably more complex and we have to entirely rely on numerical analysis. Figure \ref{fig:bestbenchmix} shows the SDP results obtained for displaced thermal states and squeezed thermal states of different purity $\mu$ \footnote{The results for single-seed POVM, which are not shown in the figure, do not coincide with the general bound for squeezed thermal states, but follow practically the same curve. For displaced thermal states, the single-seed POVM does attain the general bound. }. The truncation errors at the higher values of $r$ and $\alpha^2$ are of the order of a few percent, but the displayed values still provide a good upper bound because truncation effects tend to lower the curves. We observe that decreasing the purity has the effect of increasing the fidelity. Thus our benchmark is specially suited for test states of moderate temperature.  It is worth mentioning here that if the guess is restricted to belong to the input family (not shown in plot), the effect is the opposite: pure states provide higher fidelities than mixed states.

\emph{Acknowledgments}
We are grateful to  G. Adesso and G.~Chiribella for discussions.  We
acknowledge financial support from the Spanish MICINN, through the
Ram\'on y Cajal program (JC),  FIS2005-01369 and QOIT
(Consolider-Ingenio 2010), and from the Generalitat de Catalunya, CIRIT SGR-00185.


\end{document}